# Interacting Immediate Neighbour Interpolation for Geoscientific Data

Arya Kimiaghalam[1,3], Andrei Swidinsky[1*,2*], Mohammad Parsa[3*]

1 Department of Physics, University of Toronto, 60 St. George Street, Toronto, ON, M5S 1A7, Canada.

2 Department of Earth Sciences, University of Toronto, 22 Ursula Franklin Street, Toronto, ON, M5S 3B1, Canada.

3 Natural Resources Canada, Geological Survey of Canada, 601 Booth Street, Ottawa, ON, K1A0E8, Canada.

**Abstract** A diverse range of interpolation methods, including Kriging, spline/minimum curvature and radial basis function interpolation exist for interpolating spatially incomplete geoscientific data. Such methods use various spatial properties of the observed data to infer its local and global behaviour. In this study, we exploit the adaptability of locally interacting systems from statistical physics and develop an interpolation framework for numerical geoscientific data called Interacting Immediate Neighbour Interpolation (IINI), which solely relies on local and immediate neighbour correlations. In the IINI method, medium-to-long range correlations are constructed from the collective local interactions of grid centroids. To demonstrate the functionality and strengths of IINI, we apply our methodology to the interpolation of ground gravity, airborne magnetic and airborne radiometric datasets. We further compare the performance of IINI to conventional methods such as minimum curvature surface fitting. Results show that IINI is competitive with conventional interpolation techniques in terms of validation accuracy, while being significantly simpler in terms of algorithmic complexity and data pre-processing requirements. IINI demonstrates the broader applicability of statistical physics concepts within the field of geostatistics, highlighting their potential to enrich and expand traditional geostatistical methods.

## Author Contacts

Arya Kimiaghalam: arya.kimiaghalam@mail.utoronto.ca

Mohammad Parsa: mohammad.parsasadr@nrcan-rncan.gc.ca

Andrei Swidinsky: andrei.swidinsky@utoronto.ca

# 1 Introduction

Irregular sampling distribution and physical limitations in geoscientific survey coverage bring the need for data interpolation techniques. An extensive range of interpolation methods have been developed over the years for various types of geoscientific data. Widely used interpolation methods include Inverse Distance Weighting (IDW), Kriging, spline/minimum curvature interpolation and Radial Basis Functions (RBFs). Such methods are used in a diverse range of geophysical and geochemical survey applications such as the interpolation of soil attributes, mineral resource estimation, environmental data, geophysical data and geological surfaces (Deutsch & Rossi, 2014; Robinson & Metternicht, 2006; Briggs, 1974; Hillier et al., 2014 for a few examples of respective case studies). Most such methods use a specific set of global mathematical or statistical assumption about a dataset and use parameters to structure the interpolation around those assumptions. However, such geospatial assumptions may bring limitations to the interpolation process. For instance, IDW assumes that closer points have greater influence on the interpolated values, and that influence decreases with distance according to an inverse power law, with the power being the parameter of interpolation (Deutsch & Rossi, 2014). However, IDW is limited by its oversimplification of spatial relationships, inability to account for directional trends, sensitivity to irregular point distributions, and lack of uncertainty estimates (Burrough et al, 2015). Ordinary Kriging assumes that spatial relationships can be captured by a semivariogram, that spatial autocorrelation diminishes with distance, and that the process is second-order stationary with constant mean and variance over large enough distances (Cressie, 2015). Ordinary Kriging is limited by its reliance on an accurate variogram model, sensitivity to outliers, inability to handle non-stationary trends without modifications/pre-processing, and computational intensity for large datasets (Oliver & Webster, 2015). Spline interpolation assumes smooth variation between data points, damps and minimizes gradients for surface smoothness, and gives local control to nearby points on the surface (Briggs, 1974). Spline interpolation is limited by overfitting in noisy data, unrealistic oscillations in sparsely sampled areas, poor handling of large local gradients, and possible extrapolation significantly beyond observed data ranges (i.e., overshoots) (Lam, 1983). RBF assumes isotropic spatial variation modeled by radially symmetric functions centered on each point and governed by a specific radial function (e.g., Gaussian). RBF is challenged by the overweighing outliers in sparse data, computational expense due to dense matrix inversion, sensitivity to the choice of radial function, and the presence of anisotropy (Buhmann, 2009).

It is noteworthy that most conventional data interpolation techniques in geosciences simultaneously consider all existing measurement points for fitting and inference (Hillier et al., 2014; Oliver & Webster, 2015). However, this fact may undermine local variations in the data. A case of this occurs in minimum curvature surface fitting, where local perturbations can affect distant regions in the fit surface. On the other hand, an attempt to honour these local variations together with global fitting requirements may result in the overparameterization of the interpolation technique and negatively affect its practicality and explainability.

To offer an alternative framework, we propose a family of interpolation techniques inspired from the concept of spin models in statistical physics (Nowak, 2007) called Interacting Immediate Neighbour Interpolation (IINI) for geoscientific data that eliminate data parametrization, increase flexibility of use over different data types and significantly reduce the need for data pre-processing compared to other conventional techniques. Due to their flexibility, spin models (i.e., locally interacting constituent models) have been applied and adapted in various interdisciplinary fields such as understanding social consensus and collective decision making in sociology (Sznajd-

Weron & Sznajd, 2000), rudimentary models of biological neural networks in biology (Schneidman et al., 2006) and modeling melt pond formation on arctic ice sheets in the Earth sciences (Ma et al., 2019).

IINI techniques solely rely on local (i.e., immediate neighbour) correlations to infer missing data values, and do not assume any form of stationarity or isotropy in data, nor prescribe set degrees of smoothness for the interpolation. We present and examine two cases of IINI interpolation, involving three different datasets. The first case is the interpolation of ground gravity data from interior New Brunswick, Canada. We use this dataset to illustrate the effect of resolution and Monte Carlo parameters on IINI. We then apply IINI to the interpolation of airborne magnetic and radiometric data from central interior British Columbia, Canada, comparing our results to a minimum curvature interpolation approach.

## 2 Methodology

2.1 Gridding the Data

IINI interpolation is proposed for spatially gridded data. Therefore, for datasets that are not regular in space, a gridding procedure is required. In other words, the dataset is binned into a two-dimensional matrix, with each bin representing a pixel element in space. However, this process cannot be performed at arbitrary spatial resolutions, and the choice of gridding resolution directly affects interpolation results for most techniques (e.g., minimum curvature is highly sensitive to resolution). Therefore, the choice of grid cell size must be carefully made. For scattered data, a common practice is to consider the mean shortest distance between measurement points. An appropriate physical pixel size can be estimated using the point count density (Hengl, 2006).

$$p = c\sqrt{\frac{A}{N}} \qquad (1)$$

where A represents the total area of the survey, N represents the number of measurements, c = 0.5 for regularly distributed data points and c = 0.25 for irregular distributions. Alternatively, the spatial correlation structure of the data is used to calculate a proper pixel size, which can be chosen in the context of its variogram; to quantify the range at which data points have high pairwise correlation (Izenman, 1991). We use Eq. 1 to calculate the resolution of choice for all of the geophysical data in this study.

2.2 Definition of Dissimilarity and Lattice Energetics

Here, we demonstrate two physics-inspired means of quantifying dissimilarities between neighbouring pixels in a gridded dataset. Note that neighbouring pixels are those that have a Euclidean distance of 1 with a pixel of interest, implying that each pixel has 4 immediate neighbours (i.e., pixels located on diagonals are not considered). In addition, we elaborate on the connection and analogues between the ideas of energy in physical systems and data dissimilarity D (see Sects. 2.2.1 and 2.2.2).

*2.2.1 Square Difference*

The concept of square difference is often considered one of the most basic ways to quantify numerical differences between scalar data. Consider a square lattice configuration of pixels. For a given pixel, its weighted total square difference with its four neighbours can be expressed as the following,

$$SD(p) = \frac{\sum_{i=1}^{4} b_i (p - p_i)^2}{\sum_{i=1}^{4} b_i} \qquad (2)$$

where $p_i$ indicates the $i^{th}$ immediate neighbour of pixel p, and (for the sake of generality) $b_i$ indicating a bias factor associated with the $i^{th}$ neighbour. In the case of pixels on the grid boundaries, this sum of square differences is calculated for two (4 corner pixels) or three (pixels at the boundaries, excluding corners) neighbours. The bias parameter b can be set to indicate the importance of neighbouring training pixels. For instance, b = 1 indicates a no bias interpolation, where training and inference neighbouring pixels have equal influence over their immediate neighbourhood, and b → ∞ is the case where training pixels dictate the state of their immediately neighbouring inference pixels. There are a diverse range of approaches to assigning such weights. However, we can consider a rudimentary approach to demonstrate the use of such bias weights for IINI. For spatial coverages of less than 25%, most pixels have at most one neighbouring training pixel and; we can choose to set b = 3 for neighbouring training pixels, to give the training pixel an equal influence in the update relative to the other three neighbours combined. For a min-max normalized gridded data, the minimum square difference becomes 0 for a case of most similarity (considering no bias), and 1 for maximum difference. This function can be used as an indicator for the pixel-to-pixel dissimilarity score. Therefore, a pixel neighbouring training pixels tends to minimize its square difference with those training pixels over a sequence of updates to its value.

The square difference metric has a simple physical analogue as follows. In physical systems, the evolution of states favour a move to lower energies, often quantified by a potential energy U. A simple example of potential energy is that of a spring: $U = \frac{1}{2}k\Delta x^2$, where k is a spring constant indicating its stiffness and $\Delta x$ represents the displacement of the spring from its equilibrium position. For the use of SD(p) a clear connection emerges between minimizing the potential U and SD(p) for the IINI problem. For gridded data, we can think of each pixel being connected to its four neighbours by four fictitious springs, each being stretched by a distance equivalent to their scalar difference with their neighbour (exception being the pixels on the edges, where they have two to three immediate neighbours) (Fig. 1).

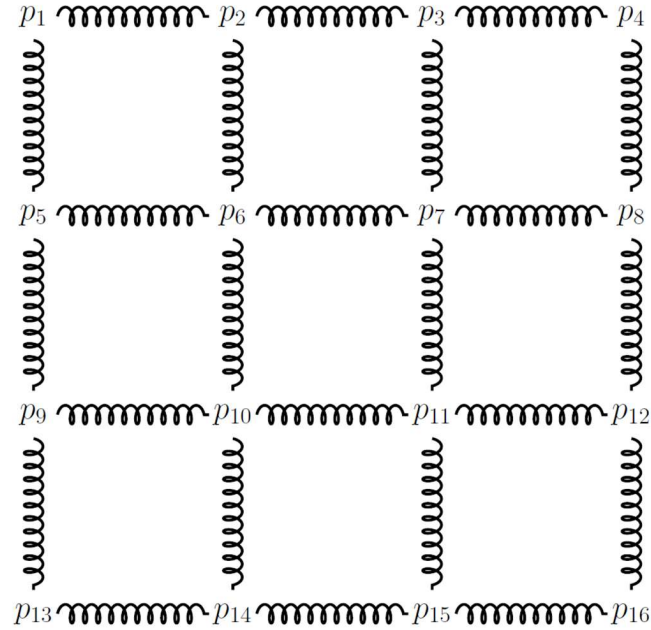

**Fig. 1** A sample of a square lattice configuration in the context of pixel-to-pixel square difference.

In this case, the total potential energy becomes the sum of potential energies from each of the four neighbours (e.g., for $p_6$ in Fig. 1, we have $U_{total}=\frac{1}{2}k((p_6-p_5)^2+(p_6-p_2)^2+(p_6-p_7)^2+(p_6-p_{10})^2)$, assuming equal spring constants). Minimizing this potential energy $U_{total}$ for every pixel altogether is equivalent to minimizing the total square difference, consistent with Eq. 2. Thus, the solution to the IINI problem is ultimately equivalent to the equilibration of a network of springs as shown in Fig. 1. Elastic spring network models are used in other scientific domains for modeling physical properties as well, such as the elastic modelling of organic and biological materials (Alzebdeh & Ostoja-Starzewaski, 1999).

*2.2.2 Cosine Dissimilarity*

Although this work solely focuses on the interpolation of scalar data, it can also be extended for the interpolation of vector data. For example, different iterations of cosine similarity metric are used for adapting magnetic field landscapes consisting of local dipole constituents in ferromagnetic material under local and external magnetic field perturbations in physics (Kosterlitz, 1974; Zittartz, 1976), which is effectively a form of dynamic interpolation. A dataset consisting of unit vectors of varying orientations can be interpolated based on a cosine similarity metric. Given a pixel and its four neighbours, the cosine dissimilarity score for its direction unit vector and that of its neighbours can be simply calculated as the negative of its cosine similarity (i.e., inner product) (Eq. 3a),

$$C(p) = -\frac{\sum_{i=1}^{4} b_i \langle \vec{p}, \vec{p_i} \rangle}{\sum_{i=1}^{4} b_i} \qquad (3a)$$

where $\vec{p}$ and $\vec{p_i}$ represent the direction unit vector of an arbitrary pixel and its $i^{th}$ immediate neighbour, with $b_i$ being the bias factor of the $i^{th}$ neighbour. As an example, the 2D case can simplify the direction component into polar coordinates, where p values are scalar angle values,

$$C(p) = -\frac{\sum_{i=1}^{4} b_i \cos(p - p_i)}{\sum_{i=1}^{4} b_i} \ . \qquad (3b)$$

For other cases involving both direction and magnitude, we propose converting the data into Cartesian coordinates and interpolate Cartesian components separately using the square difference optimization metric from Sect. 2.2.1, particularly for strongly coupled vector data such as oceanic currents and pressures (Sun, 2018). However, one can separately interpolate the direction and magnitude using the cosine and square difference metrics respectively, if these components are weakly correlated or fully independent from each other (e.g., radial force fields). In this case, a vector dataset can be divided into two datasets (i.e. polar components), a magnitude dataset and a directional dataset (i.e., unit vectors representing direction), interpolated separately, and paired together for a full representation at the end. In this case, the magnitude dataset can be interpolated using the square difference metric and the direction dataset using a cosine similarity metric.

The energy equivalent of this dissimilarity function for spins is a fundamental component in the study of spin lattice systems, an ongoing area of investigation in statistical and condensed matter physics (Kosterlitz, 1974; Zittartz, 1976). The range of angles can also be discretized for better computational efficiency. In statistical mechanics, the discretized spin variant of a generalized spin lattice system is widely known as a Potts vector model, which was developed by Renfrey Potts in the early 1950s (Potts, 1952). The Potts model plays a significant role in liquid crystal research, where it models phase transitions between crystal configurations and molecular arrangements (Bailly-Reyre & Diep, 2015). Furthermore, the model aids in studying quantum phase transitions (Rapp & Zaránd, 2006) and understanding the dynamics of interactions in ferromagnetic material (Mukamel et al., 1976).

2.3 Interacting Immediate Neighbour Interpolation (IINI)

*2.3.1 Data Pre-Processing*

We can define a dissimilarity score D for every pixel and its four immediate neighbours (see Sect. 2.2). However, in order to compute the dissimilarity scores from Eqs. 1 & 2, and commence the interpolation process, a gridded dataset with known values v is required to be normalized.

$$v \rightarrow p: \quad p = \frac{v - v_{min}^{train}}{v_{max}^{train} - v_{min}^{train}} \qquad (4)$$

A normalized data grid contains some populated and empty pixels, representing training and inference values respectively. The values of the training pixels are kept while the values of the inference pixels are randomly initialized by a discrete range of values with a spacing of $\varepsilon_p$, in other words, they are assigned random values belonging to

$$p_{choice} = \left\{ p = \left(n + \frac{1}{2}\right)\varepsilon_p \mid n \in Z^+, \left(n + \frac{1}{2}\right)\varepsilon_p < 1 \right\} \tag{5}$$

The uniform random initialization is specifically chosen because it introduces no statistical bias to the sequence of updates and the final grid configuration. While the discretization is mainly performed to save on computational costs, a recommended value of $\varepsilon_p$ can be the average error rate in the dataset. After this initialization step, each pixel is either labeled as initially empty (i.e., to be inferred) or containing known data (i.e., to be preserved during updates of inference pixels). Alternatively, they may be called inference and training pixels respectively.

*2.3.2 Monte Carlo Optimization of the Data Grid*

IINI does not directly parametrize the dataset, meaning any instance can only be described by the ensemble set of individual pixel values and their positions. Therefore, it cannot be completed using one-step optimization as it is done in least square regression or Kriging (Oliver & Webster, 2015). An effective family of algorithms in tackling complex optimization problems are Monte Carlo methods, which are highly effective at solving such problems due to their ability to explore large and complex search spaces. Traditional optimization techniques often get stuck in local minima, especially in non-linear or highly multivariate functions. The Monte Carlo framework, by contrast, uses random sampling to explore a wide range of possible solutions and configurations, which allows them to overcome local traps and move toward the global optimum. The framework can also be further modified to deal with especially challenging optimization landscapes such as uniqueness issues in protein folding (Li & Scheraga, 1987). Their stochastic nature makes them adaptable, robust, and well-suited for handling high-dimensional and complex optimizations (Li & Scheraga, 1987; Newman & Barkema, 2023; Robert & Casella, 2004). In this work, we iteratively update the pixel grid through a Monte Carlo algorithm to approach the optimal data configuration and consider the sequence of updates to be a Markov chain (Robert & Casella, 2004). This implies that each update to the inference pixels for dissimilarity optimization relies solely on the current state of the data grid, with no dependence on prior configurations. For iteratively optimizing D values of inference pixels, we employ the Metropolis algorithm where for every update, a random inference spin is chosen and its initial dissimilarity calculated. A pixel value is randomly (i.e., from a flat probability distribution) chosen from the set $p_{choice}$ (Eq. 5) as a potential update candidate, and dissimilarity is re-calculated for the update candidate. The Metropolis algorithm accepts any update that decrease dissimilarity while only accepting mistakes with a $exp(-\Delta D/T)$ probability, where T is a constant, indicating tolerance towards suboptimal updates. If an update is accepted by the algorithm, the inference pixel value is replaced with the new value and left unchanged otherwise. Other Markov chain Monte Carlo methods such as Gibbs sampling, are successfully used for image reconfiguration and moderate noise reduction, under the assumption of known image distributions (Geman & Geman, 1984).

Hypothetically, this algorithm may continue to run indefinitely, however, a practical stopping criterion may be designed to accompany the algorithm to further improve computational efficiency.

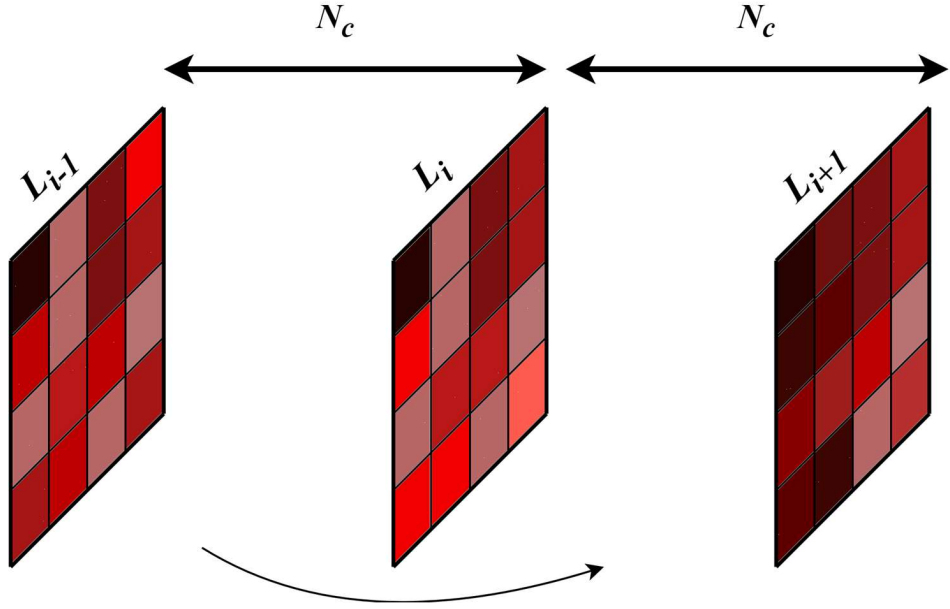

**Fig. 2** An intermediate step in the stopping mechanism of the Metropolis algorithm. The RMSE between two consecutive lattice checkpoints is calculated and the continuation of updates is conditioned upon RMSE values satisfying RMSE($L_i$, $L_{i-1}$) > 0.5.

To design the stopping criterion for the Metropolis algorithm, we rely on the discretization error $\varepsilon_p$ and a series of checkpoints for the intermediately updated grids. These checkpoints are placed at equal iteration count separations from each other. The goal is to give every inference pixel of the data grid enough opportunity to experience all possible updates in the set $p_{choice}$ over updates, which contains roughly $1/\varepsilon_p$ values. Therefore, given $N_{infer}$ number of inference pixels on a data grid, and that inference pixel locations are randomly chosen for updates, the checkpoints need to be $N_c = N_{infer}/\varepsilon_p$ iterations apart (see Fig. 3). For example, if we have a 50 by 50 grid of data with 75% empty cells and a 2.5% discretization, the grid checkpoints are set 75000 updates apart. As more Monte Carlo steps (i.e., pixel updates) are taken, the grid at every checkpoint is compared to that from the previous checkpoint through a relative Root Mean Square Error (RMSE) calculation. It is expected for these RMSE errors to continue decreasing as the grid approaches an optimal grid configuration. A grid checkpoint becomes the stopping point if its RMSE with the previous grid checkpoint become less than $\varepsilon_p/2$. This implies that the interpolation process stops when the expected change/update of inference pixels over two consecutive checkpoints is reduced to half of the discretization interval.

In the early steps of the Monte Carlo process, the updates to the pixels are highly susceptible to the random initialization of the inference pixels. While the probabilistic acceptance of apparent mistakes is incorporated into the Metropolis algorithm to prevent trapping in local minima of $D_{total}$, the probability of accepting mistakes of any magnitude must dynamically diminish over Monte Carlo iterations. If not, the Metropolis algorithm keeps accepting mistakes in much later stages, which would significantly slow down the interpolation process. On the other hand, if all mistakes are rejected at all steps of the process, the interpolated grid configuration would not represent the true optima of the D landscape and only show a local optimum. Ultimately, rejecting all apparent mistakes is what we desire, at which point the influence of random pixel initialization is largely removed. To strike a balance between these two competing aspects, we may gradually decrease the Metropolis mistake acceptance rate from an initial high to zero over the Monte Carlo steps by varying the parameter T in the acceptance rate expression. This process is referred to as simulated annealing in condensed matter physics (Geman & Geman, 1984, Van Laarhoven et al., 1987). Higher T values increase the mistake acceptance rate while the contrary (i.e., as T→0) leads to rejection of all mistakes. A conservative approach is to start with a 50% acceptance rate for the worst possible mistake (i.e., ΔD=1) at first. Thus, the starting T can be calculated as $T_{start} = 1/ln(2)$. Note that this choice of $T_{start}$ is a generally recommended value, which may be changed to lower values depending on the complexity of the dataset or processing time constraints by the user (e.g., a careful choice of smaller $T_{start}$ reduces computing time).

To let T approach zero, the initial temperature $T_{start}$ can be decayed exponentially over checkpoints such as $T(n) = T_{start}/a^n$, where n is the number of the checkpoint, and a is a decay constant. However, to reach the best grid configuration, the Monte Carlo process becomes progressively slower as the frequency of acceptable updates is greatly diminished. Therefore, we employ an analytical value approximation algorithm to bridge the gap between the final Monte Carlo results, and the optimal grid configuration.

*2.3.3 Analytical Value Approximation*

After the Monte Carlo updates are completed, the resulting grid is adequately close to the most optimal grid configuration since the initial randomness of the grid configuration is suppressed, and longer-range correlations are consolidated. This suggests that every pixel in the data grid is greatly optimized relative to its immediate neighbours (and similarly for the neighbouring pixels). Thus, for a given pixel, we can lock its neighbours and directly update pixels by analytically solving for the optimal pixel value; one that produces the highest immediate neighbour similarity. For the two metrics presented in Sect. 2.2, the analytical solution is calculated to be the solution of a first order optimization. For square difference dissimilarity, we have

$$\frac{d(SD(p))}{dp} = \frac{2\sum_{i=1}^{4} b_i(p-p_i)}{\sum_{i=1}^{4} b_i} = 0,$$

$$p_{optimal} = \frac{b_1 p_1 + b_2 p_2 + b_3 p_3 + b_4 p_4}{\sum_{i=1}^{4} b_i}. \tag{6}$$

And for reference, in the case of cosine dissimilarity (2D example) we have

$$\frac{d(C(p))}{dp} = \frac{-\sum_{i=1}^{4} -b_i \sin[(p-p_i)]}{\sum_{i=1}^{4} b_i} = 0,$$

$$\sum_{i=1}^{4} b_i \sin(p)\cos(p_i) = \sum_{i=1}^{4} b_i \cos(p)\sin(p_i),$$

$$\tan(p) = \frac{\sum_{i=1}^{4} b_i \sin(p_i)}{\sum_{i=1}^{4} b_i \cos(p_i)},$$

$$p_{optimal} = \tan^{-1}\left[\frac{\sum_{i=1}^{4} b_i \sin(p_i)}{\sum_{i=1}^{4} b_i \cos(p_i)}\right]. \tag{7}$$

Note that unlike the Monte Carlo updates, the updates in this analytical step are no longer discrete. This analytical value approximation process is repeated in multiple rounds over all inference pixels and stopped if a desired convergence is reached and can act as an effective shortcut to the Monte Carlo optimization of the data grid. It can also be applied earlier than the designated stopping criterion by the user, to further improve processing time at the cost of interpolation quality.

For conditional interpolation, the training pixel values are preserved while the inference pixels are analytically approximated as described previously. However, for an unconditional interpolation, the training pixels may also be replaced by their analytically approximated values once the inference pixels are properly calculated and locked in. Note that this process is identical to the analytical approximation of inference pixels, only applied to training pixels. The re-calculation of training pixels may implicitly help with improving smoothness. Note that this approach may slightly reduce the range of the interpolated dataset. As the range of inference pixels is theoretically a subset of the training pixels, any recalculated training pixel values are naturally confined to the limits defined by the inference pixel range. This narrowing becomes less noticeable as the grid cell size is reduced, which makes the transition between training pixels more resilient.

## 3 Effect of Grid Resolution, Simulated Annealing and Assigned Bias on IINI

In this section, we showcase the effects of varying IINI parameters and spatial resolution of data on the results of IINI interpolation. We use gravity data catalogued by the Geological Survey of Canada (Geological Survey of Canada, 2018) to illustrate the effects on real world data.

3.1 Effect of Spatial Resolution on IINI Interpolation

Proper gridding of irregularly distributed datasets is critical to the interpolation process. The method and logic for gridding can vary based on the data being used (Hengl, 2006). Here, we gridded and interpolated irregularly distributed gravity data that span an area of roughly 30km by 30km in size, from interior New Brunswick, Canada. The data is gridded into the resolutions of 100m, 200m and 400m, and separately interpolated with the following parameters (a=1.15, b=1, $\varepsilon_p$ = 1/50), the results of which are illustrated in Fig. 3a. (Note that the optimal grid size is 200m according to Eq. 1).

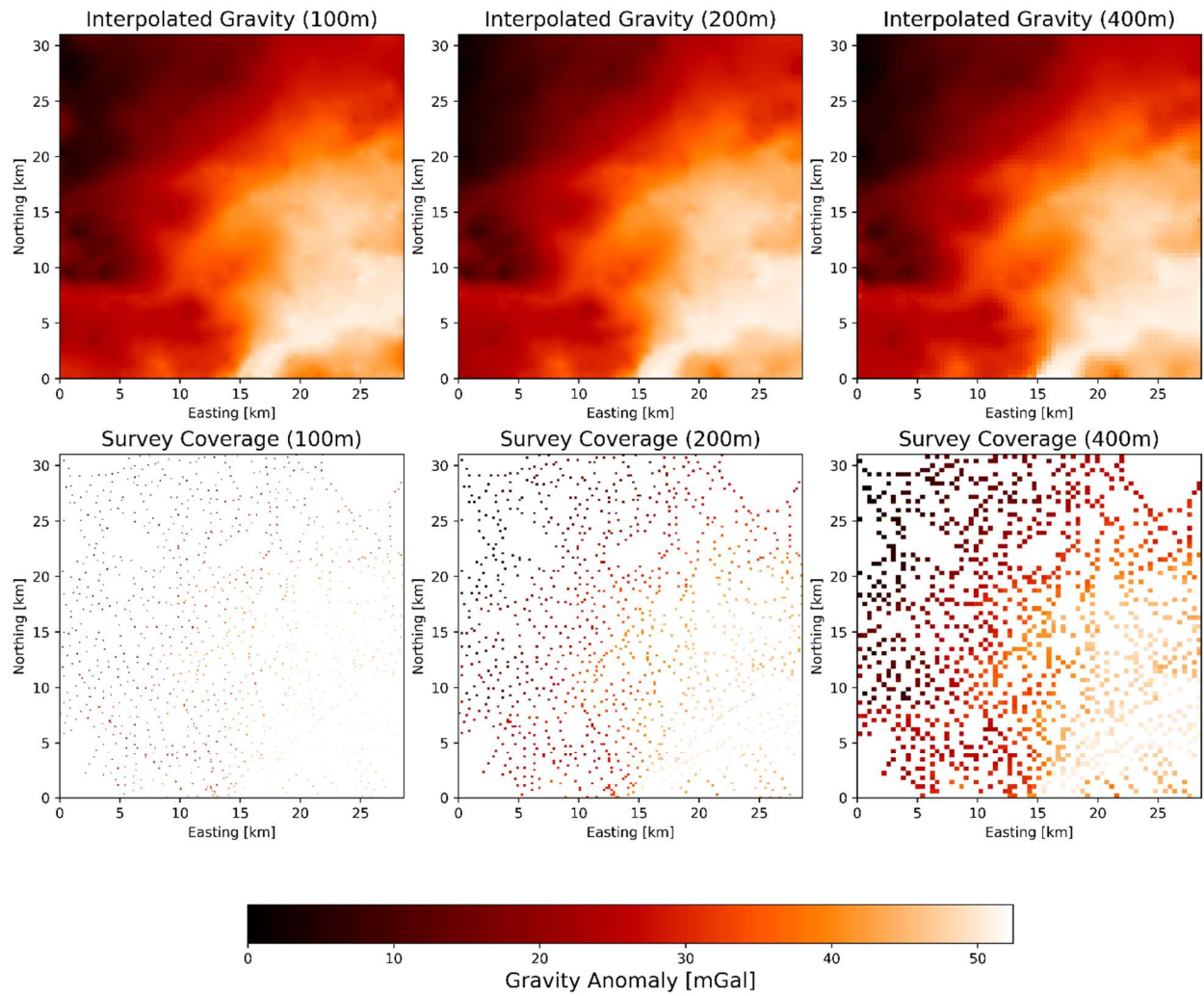

**Fig. 3a** Interpolated gravity data, in three different grid resolutions (100m, 200m, 400m) (top row). Coverage of known data at different grid resolutions (bottom row).

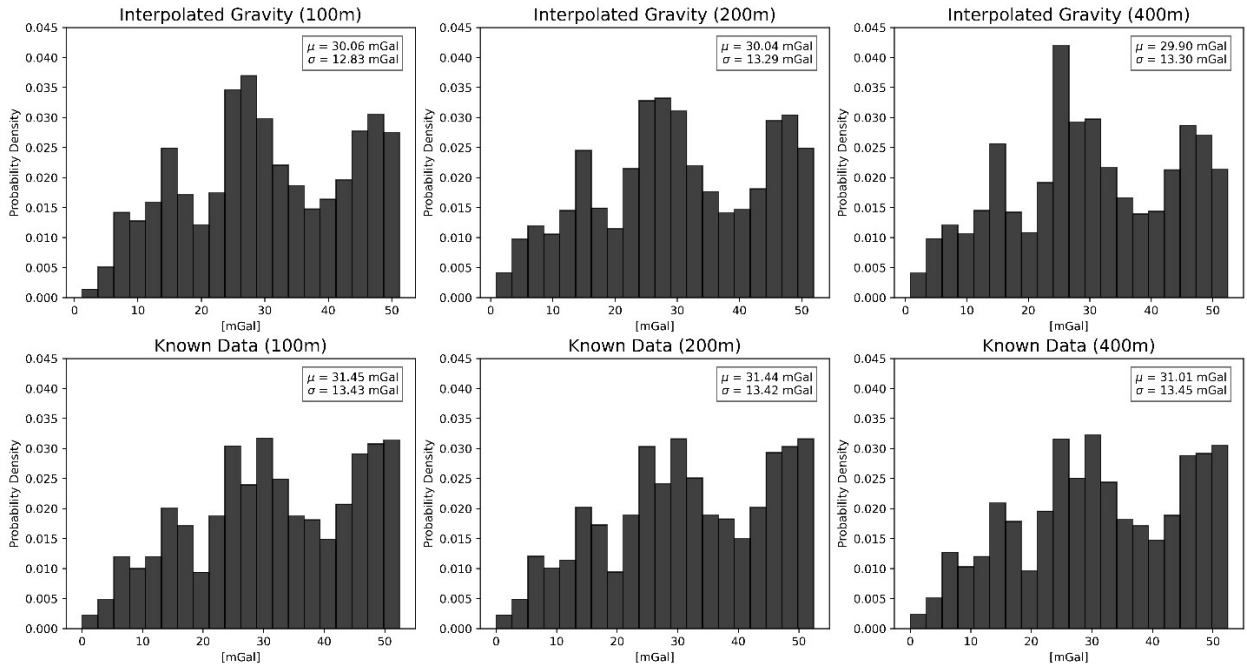

**Fig. 3b** Interpolated gravity data histograms, in three different grid resolutions (100m, 200m, 400m) (top row). Histograms of known data (i.e., training data) at different grid resolutions (bottom row). Note that the interpolated gravity histograms include the re-interpolated training pixels.

Three observations can be made from the histograms as well as directly from the interpolated maps. First, regardless of resolution, interpolation results consistently produce less extreme gravity values compared to that of the training data histogram (see Fig. 3b), although other parts of the histogram match. Second, the standard deviation of the interpolated values decreases with a drop in grid size due to mean reversion. This result is connected to our third observation, which is that IINI interpolation becomes spotty-looking and isolates large-gradient anomalies under a low grid size setting, leading to mean reversion around these anomalies (this is particularly evident in low gravity anomalies in the center-East area of the region in Fig. 3a). This can be attributed to the fact that IINI relies on pixel-to-pixel information propagation and assumes a discrete space to calculate interpolations. When the grid size is reduced, the physical distance between pixel centroids remains the same, however, their separation in terms of pixels increases significantly. This is a limitation of IINI compared to methods that use continuous surfaces for interpolation, such as Kriging and minimum curvature surfaces, which calculate their estimates on a continuous space before projecting them back onto a grid of choice. By placing too many pixels between inference and training pixels, their ability to communicate through local pixel-to-pixel interactions is reduced, isolating them. The resulting mean reversion can be identified in the respective interpolated value histogram of the gravity map with a 100m grid size. Therefore, the choice of the proper resolution is key for practicality. Though, it should be noted that this problem is absent for the grid size that was properly calculated according to the survey density and extent (i.e. 200m), implying that this concern is practically avoided by a careful choice of grid size.

3.2 Effect of Rapid Annealing on IINI Interpolation

Here we fix the grid resolution to 200m, the bias factor to $b = 1$, and use two choices of "a" ($a = 1.15$ and $a = 10^5$ for our annealing parameter. This means that in the case of the high annealing parameter, the IINI algorithm effectively does not tolerate any suboptimal update beyond the first few checkpoints, while in the case of slow annealing, the algorithm gradually becomes more selective over checkpoints. Fig. 4 shows the results of both interpolation results, together with a difference map.

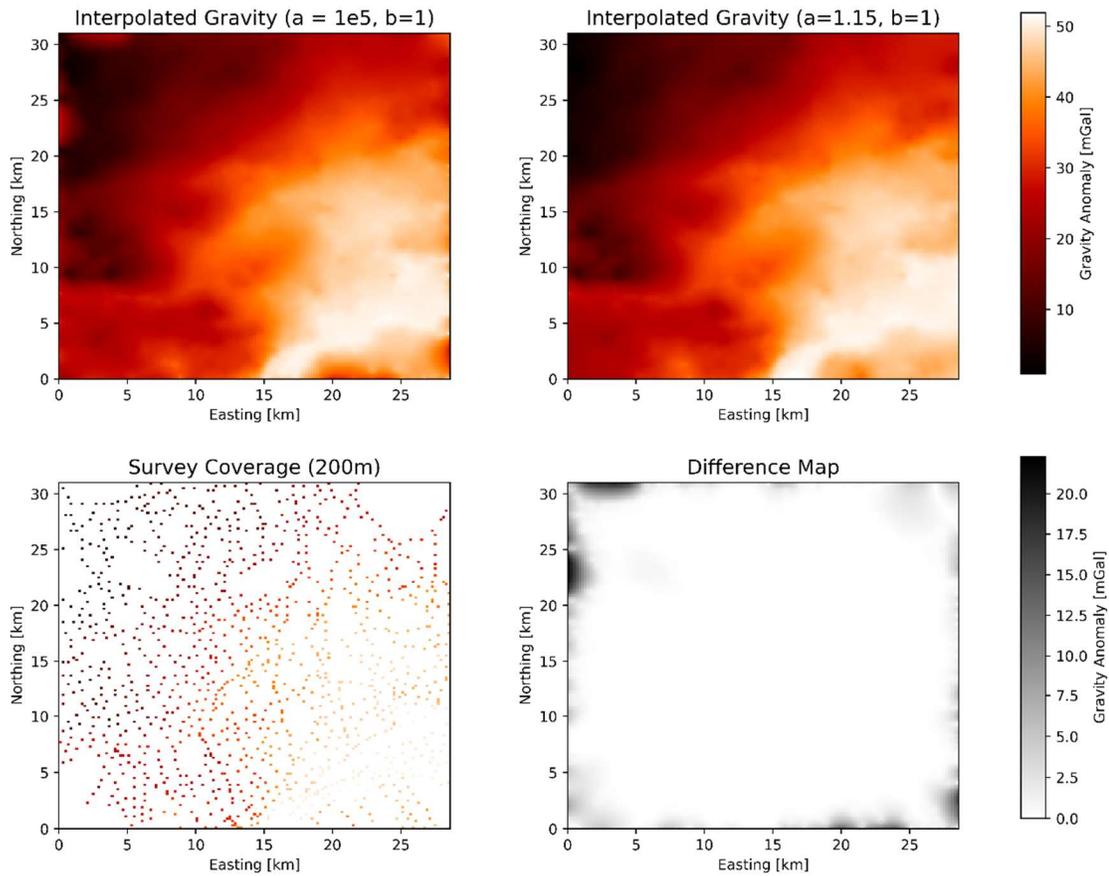

**Fig. 4** Interpolated gravity data for a = $10^5$ (top left), interpolated gravity data for a = 1.15 (top right), known data pixels (bottom left) and a difference map between the extreme and moderate annealing cases (lower right).

The interpolation result produced from a high annealing parameter exhibits similar outcomes over the well-covered areas of the map. However, the discrepancy arises in the areas that are not as well covered by measurements. Some clear examples include the South-West, North-East, and North-West corners of the survey area, which are highlighted in the difference map. In these regions, inference pixels are far from any training pixel and are less constrained. This lack of constraint and relatively high distance (counted in pixels) from training pixels causes a reversion to the mean in the rapid annealing case, where pixels do not find the opportunity to effectively communicate with distant training pixels. Slow annealing (with a value of 1.15 in this case) enables the effective propagation of information from distant training pixels, while also providing stronger constraints on the optimal configuration of inference pixels in baren regions. In contrast, rapid annealing (i.e., high values of "a") makes the interpolation prone to regressing to the mean in baren and isolated areas of the data.

3.3 Effect of Biased Weighing of Pixels on IINI Interpolation

In this section, we employ our rudimentary bias assignment approach outlined in section 2.2.1, and investigate how it changes the statistics of the interpolated gravity data. Note that we use a resolution of 200m and an annealing parameter of a = 1.15 for all interpolations in this section. Fig. 5 illustrates the interpolation results for both biased (b = 3 for training pixels) and unbiased (b = 1 for training pixels) weighing of pixels. Note that for the data illustrated in Fig. 5, 99% of inference pixels have one training pixel neighbour at most, which supports our assumption in Sect. 2.2.1.

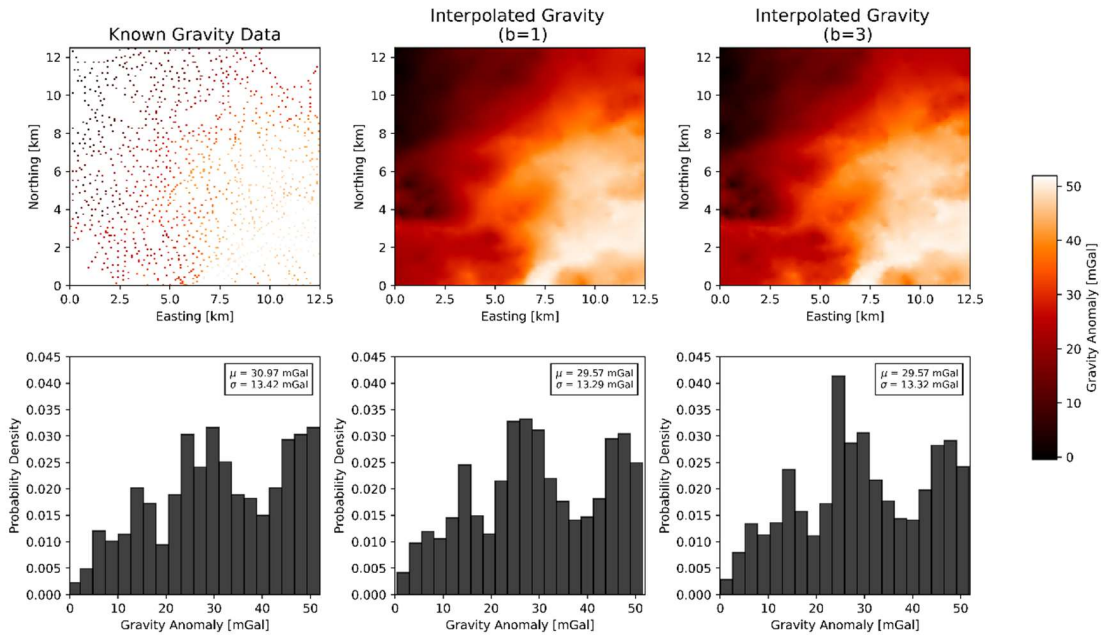

**Fig. 5** Interpolated gravity data through a biased (b = 3 for training pixels) and unbiased (b = 1 for training pixels) pixel weighing approach (top row). Histograms of both interpolation results and the gravity measurements (bottom row), ($\varepsilon_p$ = 1/50). Note that the interpolated gravity histograms include the re-interpolated training pixels.

We can make three important observations concerning these results. The first observation, which both biased and unbiased IINI results share, is a drop in the abundance of pixel values in the extrema (extreme high and low gravity measurements), and a minor contraction in the range of data values, similar to the results in Sect. 3.1 (see Fig. 5 for reference). The second observation is the pronounced difference of the biased IINI histogram around the mean compared to the unbiased and training data cases. However, the histogram of interpolated values and that of the original data are highly similar in other parts of the histogram. The standard deviation for both the biased and unbiased weighing results are close to identical, however, they are slightly smaller (by ~1%) compared to that of the known data histogram. The third, and most visual observation is the "blockiness" of the biased IINI interpolation maps, as opposed to smoother features in the unbiased

IINI interpolation. This is mainly due to the way that weights are assigned, which heavily favour training pixel neighbours (i.e. known data) and bolden their influence only in their immediate neighbourhood and not beyond. We shall see similar effects in Sect. 4, where we interpolate magnetic and radiometric data.

## 4 Interpolation of Airborne Radiometric and Magnetic Data

4.1 Data Description

The airborne radiometric and magnetic data were collected using a helicopter-borne platform by Fugro Airborne Surveys, at the request of the Geological Survey of Canada in 2004 (Geological Survey of Canada, 2004). The survey region is a 12.5km by 12.5km crop of land encompassing Helene Lake and Hannay Lake in central British Columbia, with an average flight line separation of 250m. To interpolate the gaps between flight paths, we first gridded the dataset to a 50m resolution. This choice of resolution ensures that the gridded flight paths remain connected or shifted by at most one pixel across the paths. It also closely matches our proper resolution estimate of 52m from Eq. 1 as well as that of the suggested resolution provided by Fugro Airborne Surveys (50m).

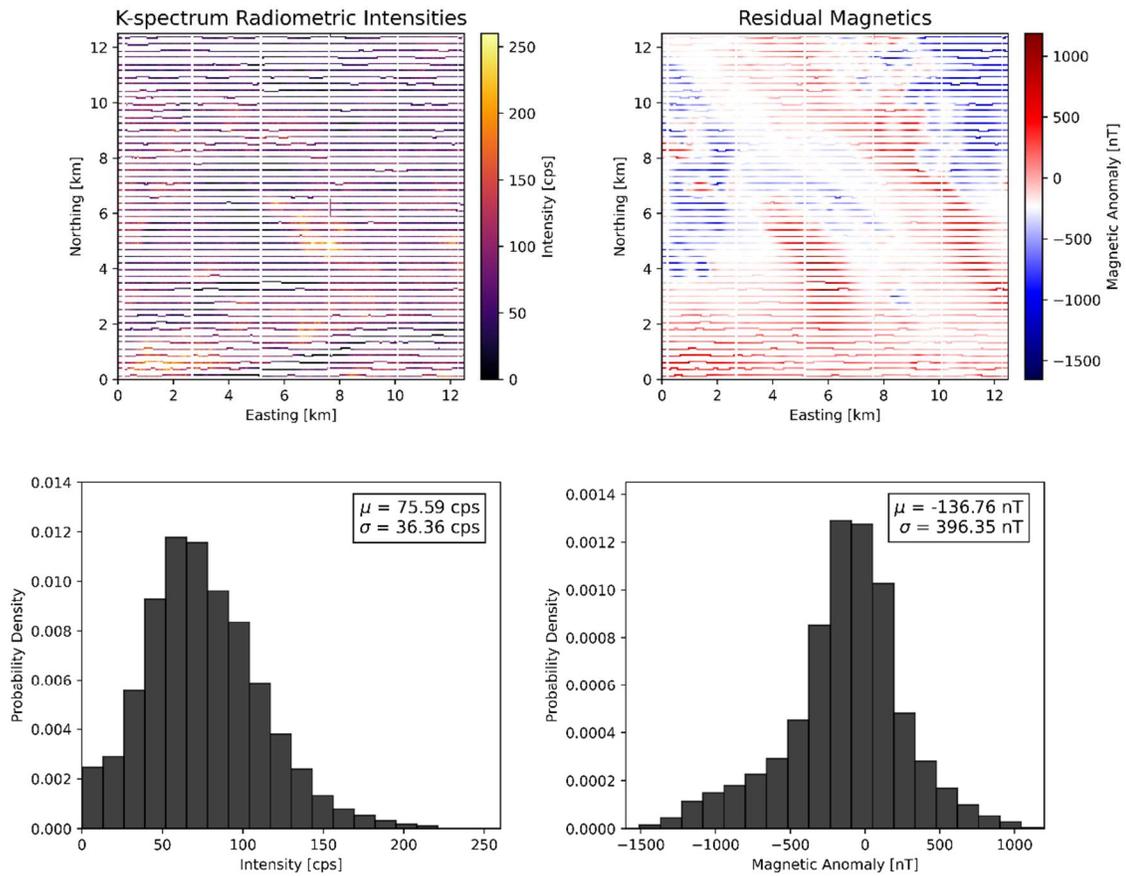

**Fig. 6** K-spectrum radiometric and residual magnetics data taken from an airborne survey performed in the proximity of Hannay and Helene Lakes, British Columbia, Canada. The gridding size of the raw data is chosen to be 20% of the average flight line separation (50 meters).

The choice of gridding resolution leads to a 23% spatial coverage. In addition, there are 6 tie lines in the dataset, which will serve as validation points for our interpolation results.

4.2 Interpolation Results

We interpolated this data using the square difference formulation of IINI (both with biased and unbiased weighing of training pixels), as well as minimum curvature interpolation for comparison (note that no damping was applied to the minimum curvature approach). The python package **Verde** is used to perform minimum curvature interpolation (Uieda, 2018). For the radiometric data, the minimum curvature interpolation yielded a Root Mean Square Error (RMSE) of 16.2cps, while IINI yielded an RMSE of 14.34cps and 14.30cps for the unbiased and biased iterations respectively. For the magnetic data, the minimum curvature approach produced an RMSE of 44.94nT while the biased and unbiased iterations of IINI produced an RMSE of 44.63nT and 51.11nT respectively. IINI and minimum curvature results are shown in Fig. 7 & 8. The evolution of the IINI interpolation grid is also illustrated in Fig. 9, which illustrates the acceleration of convergence to more optimal pixel values at mid-range checkpoints.

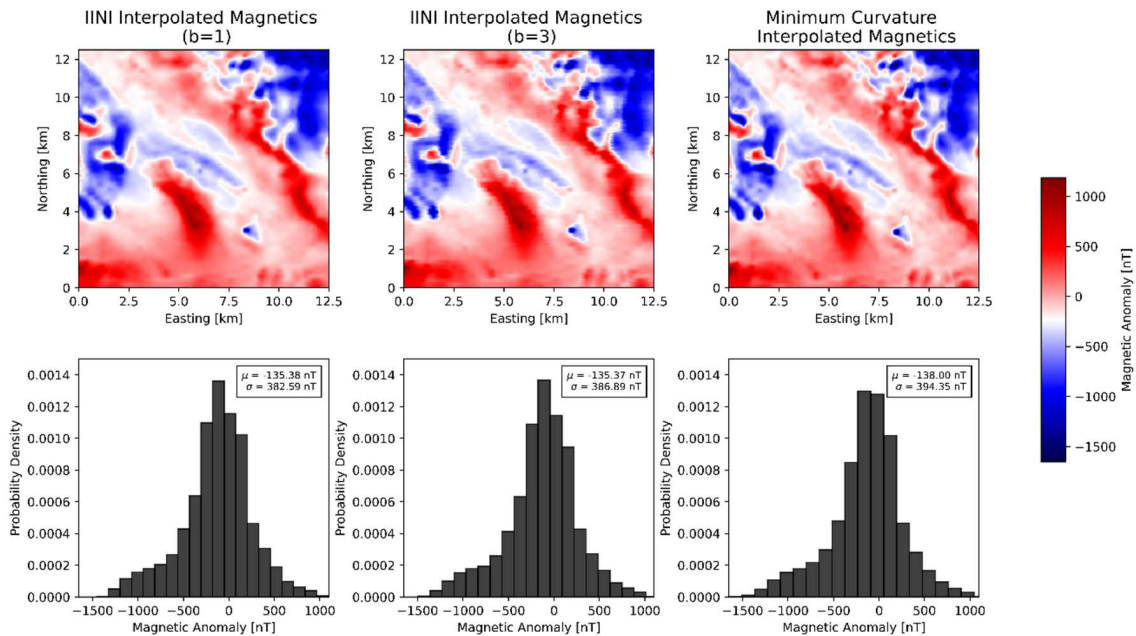

**Fig. 7** Interpolation of magnetics data using a biased and unbiased IINI as well as minimum curvature results for comparison (top row), histogram of interpolated values (bottom row). IINI interpolation is performed with (a = 1.15, $\varepsilon_p$ = 1/50). Note that the interpolated magnetic histograms include the re-interpolated training pixels.

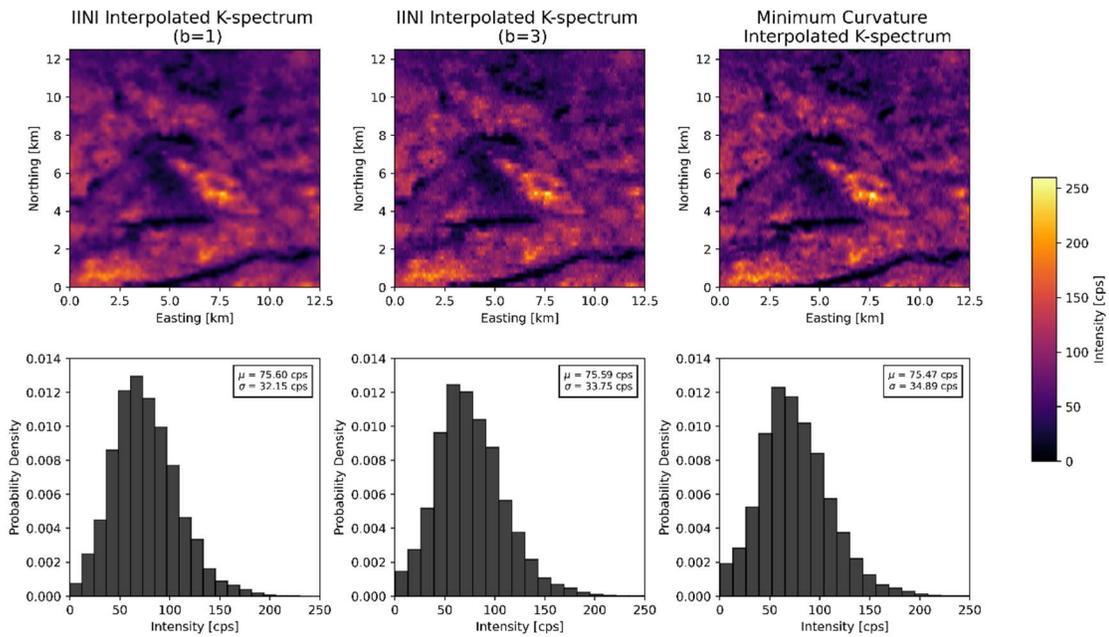

**Fig. 8** Interpolation of airborne K-spectrum radiometric data using a biased and unbiased IINI as well as minimum curvature results for comparison (top row), histogram of interpolated values (bottom row). IINI interpolation is performed with (a = 1.15, $\varepsilon_p$ = 1/50). Note that the interpolated radiometric histograms include the re-interpolated training pixels.

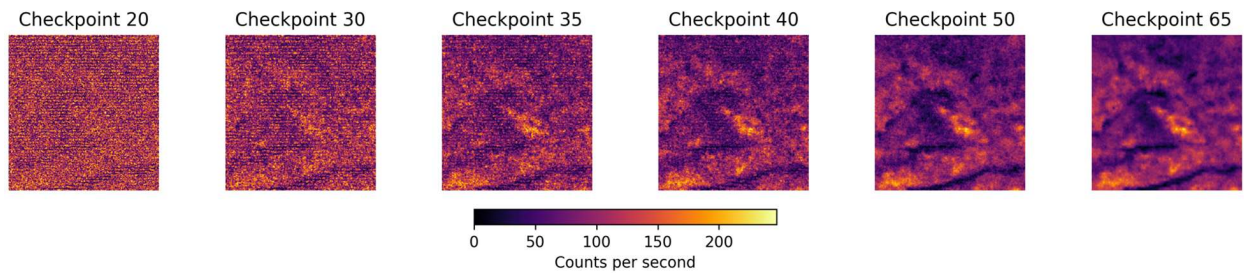

**Fig. 9** Evolution of the IINI data configuration of the airborne radiometric data in Fig. 8 at different checkpoints.

A prominent distinguishing feature of the IINI interpolation result is its relative smoothness compared to that of minimum curvature in this application. The minimum curvature interpolation shows more sharpness around anomalies and contains a sharper transition between lows and highs, particularly in interpolated radiometric data. However, this comes at the cost of returning extreme values during the interpolation, as it is challenging for minimum curvature approaches to deal with large local gradients. For instance, if there is an extreme high to extreme low transition over a short distance, minimum curvature attempts to capture the large gradient between these points. Since by its design, such methods attempt to fit the data using a minimally curved surface, this capture of large local gradients leads to large overshoots to extreme values away in these regions; before the surface rebounds to fit the rest of the data (Lam, 1983). For the radiometric data in this study, this phenomenon occurs at the boundaries of Hannay and Helene lakes (seen as dark continuous shapes in Fig. 8), with the overshoot happening near the midpoints of the lakes (low count regions). For such cases, the interpolation results are capped, or the minimum curvature damping factor is sufficiently increased to lower its susceptibility to extreme gradients, which can affect all data points and significantly reduce the overall sharpness of the interpolated data (the minimum curvature results in Fig. 7&8 were manually capped at the extrema of the dataset after interpolation). IINI faces no such challenges, since its initialization and normalization steps ensure bounded interpolation results.

## 5 Discussion

We have seen IINI used for the interpolation of ground and airborne geophysical data (gravity, magnetics and radiometric surveys). In data reconstruction, different iterations of IINI showed competitive validation performances relative to conventional interpolation methods at relatively low data coverage rates (< 25%). We also showed the effect of varying grid resolution, known neighbour biases and the pace of simulated of our Metropolis algorithm on the results of IINI interpolation and its statistics. We observed that slow annealing and a proper choice for grid resolution represent the most influential controllable factors in IINI interpolation (note that in principle, the bias assignment is not required for IINI).

Furthermore, the processing time of IINI can be improved by the choice of a smaller $T_{start}$. Decreasing $T_{start}$ by a factor of 10 from its conservative estimate significantly improved the processing time, while it slightly decreased the validation accuracy and mean reversion as described in Sect. 3.2. Therefore, the user is free to reduce $T_{start}$ from the general recommendation (see Sect. 2.3.2) given the trade off in validation accuracy is acceptable. However, this choice is highly data-dependent, and no rule can be made for all cases.

Since IINI relies on immediate neighbour similarities, it can bring unique advantages to some interpolation cases. For example, consider an ideal gridded airborne geophysical data, with parallel and continuous (i.e., unshifted) flight lines. In principle, IINI defines immediate neighbors as those separated by a Euclidean distance of 1. Consequently, inference pixels on either side of any training pixel along the flight lines cannot interact. This is because the training pixel, preserved during the Monte Carlo phase of IINI, acts as a barrier, preventing any mutual influence between neighboring inference pixel segments. Therefore, the entire survey region can be segmented by the flight lines, with each segment containing two flight lines and the gap in-between. Since no pixel in a segment can interact with pixels in other segments, each segment can be interpolated independently – a procedure which can be easily parallelized on multicore processors to reduce computing time. To illustrate the potential of this strategy, we can consider the airborne data in this study. It contains 52 flight lines, implying that there are 51 segments in between flight lines. Assuming continuity of flight lines, the interpolation process can become 51 times faster just by simple parallelization, further showcasing the inherent adaptability of IINI.

The second practical feature of IINI, in theory, is its adaptability to any well-defined graph (Bondy & Murty, 1976) or other regularly structured datasets beyond square grids. For example, IINI can easily be adapted to honeycomb lattices and their corresponding data frames by modifying

Eqs. 2 and 3 such that each data point now has six immediate neighbours. Such flexibility is particularly useful in reducing pre-processing and regularization of data histograms for geoscientific data stored in a non-square grid format.

The third feature of IINI is its inherent suppression of local noise and extreme gradients. Since IINI is only reliant on local similarity of measurements, individual pixels are not able to directly project their influence beyond their immediate neighbours. Therefore, if local pixels collectively defy a noisy pixel, their hegemony prevents the influence of the noisy measurement from spreading significantly, reducing the susceptibility of IINI to local noise.

IINI principles may also face challenges, such as the case of extreme data discontinuity. The immediate neighbour similarity principle in IINI can potentially put an upper cap on the magnitude of local gradients, since transitions between a local pair of training pixels are highly constrained by their relative distance in pixels. This effect often manifests itself as a sharp drop from a known local anomaly (i.e., a training pixel containing an extreme value within the data range), and usually occurs due to noise or the use of improper grid resolution. A second challenge is a tendency to mean revert in distant barren parts of a data grid. Since there are no training pixels to convey local information over a practical number of IINI iterations, the inference pixels tend to largely align themselves in pockets of similar values, most of which assume the expected value of the set $\langle p_{choice} \rangle$ to be 0.5. This is expected, as IINI consolidates long-range correlations through the initial development of local correlations (i.e., similarities) rather than empirically determining them through the calculation of a semivariogram.

## 6 Conclusion

This study proposes a family of interpolation techniques for geoscientific maps based on immediate neighbour similarities, inspired from spin systems in statistical physics. Two formulations of immediate neighbour similarity are presented along with their direct and tangible analogues in classical physics. The proposed interpolation techniques show a competitive performance in comparison to established interpolation techniques in the geosciences (e.g., minimum curvature) and are shown to have unique advantages in terms of versatility, lack of global parametrization, computational efficiency, parallelization and algorithmic simplicity.

The flexibility and non-parametric nature of IINI can potentially be exploited in future work for the augmentation of 2D geoscientific data in machine learning applications, particularly in cases where proper training data availability is limited. Data augmentation can involve randomly removing portions of the data and reconstructing them through IINI interpolation. In this situation, the non-parametric and unbiased nature of IINI can make machine learning models less susceptible to interpolation-related prediction biases during training with augmented data.

## Acknowledgement

We would like to thank Dr. Renato Cumani from Natural Resources Canada, for offering their valuable insight during manuscript preparation, as well as Peter Tschirhart and Dr. Kamran Esmaili for helpful discussions. Arya Kimiaghalam acknowledge the support of the Critical Mineral and Geoscience Data (CMGD) Program of the Geological Survey of Canada for providing funding for this research.

# Declaration

## Funding

This research was supported by the Critical Mineral and Geoscience Data (CMGD) Program of the Geological Survey of Canada, Natural Resources Canada.

## Conflicts of interest/Competing interests

Not applicable.

## Availability of data and materials

All data used in this research is public and cited in this manuscript.

## Code Availability

The code will be made available upon request for research or academic purposes.